\documentclass[prb,aps,twocolumn,superscriptaddress]{revtex4}
\usepackage{graphicx,amsmath,bm,xcolor}
\usepackage{dcolumn}

\begin{document}

\title{
Hidden Bose-Einstein Singularities of Correlated Electron Systems:\\
 III. Thermodynamic Signals
}

\author{Takafumi Kita}
\affiliation{Department of Physics, Hokkaido University, Sapporo 060-0810, Japan}
\date{\today}

\begin{abstract}
We study thermodynamic consequences of the hidden Bose-Einstein singularities, which have been predicted to 
cause a pseudogap phase, based on quantum field theory of ordered phases.
Starting from the Luttinger-Ward functional for the grand thermodynamic potential,
we derive expressions of the Helmholtz free energy, internal energy, entropy, and heat capacity
in a form suitable for numerical studies. 
They are applied to the weakly attractive Hubbard model in three dimensions to calculate the thermodynamic potentials
numerically and continuously for the normal, pseudogap, and superconducting phases on the same footing.
It is shown that the entry into the pseudogap phase is detectable as singularities of thermodynamic potentials, especially a discontinuity in the heat capacity.
\end{abstract}

\maketitle

\section{Introduction}

It was shown previously\cite{Kita24,Kita25} that there may exist a novel kind of singularities
in correlated electron systems called {\it hidden Bose-Einstein (BE) singularities}
which originate intrinsically from correlations.
They possibly lie in the zero Matsubara-frequency branch of the logarithmic 
correlation contribution to the grand thermodynamic potential; they are
defined by the point at which the argument of the logarithm hits zero.
It is predicted that the BE singularities are reachable independently without accompanying any manifest broken symmetries,
characterized by the emergence of the unusual {\it one-particle-reducible (1PR)} structure in the self-energy,
and signaled by the pseudogap behavior in the single-particle density of states.
Especially, it has been shown that there should be a pseudogap phase above the superfluid phase
in the weakly attractive Hubbard model.
On the basis of these results,\cite{Kita24,Kita25} 
we here investigate how the singularities affect thermodynamic potentials based on
quantum field theory of ordered phases\cite{Kita11} starting from the Luttinger-Ward (LW) functional
for the grand thermodynamic potential.\cite{LW60}

Quantum field theory has proved extremely useful for 
describing properties of many-particle systems,
as can be seen by looking into classic textbooks.\cite{KB62,AGD63,FW71,NO88,Mahan00}
However, the main focus there is on Green's functions
that are relevant to one- and two-particle properties rather than thermodynamic potentials.
Moreover, ordered phases such as superconductivity are treated almost exclusively
within the mean-field framework.
Indeed, there seems to have been few detailed studies on how to calculate
thermodynamic potentials of normal and ordered phases with correlations 
microscopically and continuously through phase transitions.
One of the exceptional ones was given by Rainer and Serene\cite{RS76,SR83} on thermodynamics of 
superfluid $^3$He, where correlation effects are treated perturbatively as strong-coupling effects 
on the mean-field results by using the LW functional.
Another study to be mentioned was given by Haussmann {\it et al}.\ \cite{HRCZ07}
on thermodynamics of the BCS-BEC crossover of atomic gases based on the attractive Hubbard model,
who adopted the De Dominicis-Martin (DDM) formalism\cite{DDM64} of expressing $\Omega$ as a functional of Green's functions and two-particle vertices in addition to the LW framework. 
However, (i) their mixed use of the DDM and LW formalisms and (ii) the renormalization of the bare interaction vertex obscure the exact procedure to obtain the thermodynamic potentials.

With these backgrounds, the aims of the present paper are twofold. 
First, we establish how to calculate thermodynamic potentials with correlations numerically with high precision
based on the LW functional alone.
Using them, we perform a numerical study on the Helmholtz free energy ${\cal F}$, internal energy ${\cal E}$, entropy ${\cal S}$, and heat capacity ${\cal C}_V$
of the weakly attractive Hubbard model investigated previously\cite{Kita24,Kita25} 
to clarify how its hidden BE singularity is detectable thermodynamically.
It will be shown that the entry into the pseudogap phase with the 1PR structure in the self-energy
is distinguishable by cusps in ${\cal E}$ and ${\cal S}$ and discontinuity in ${\cal C}_V$, 
similarly as those of the mean-field second-order transitions with broken symmetries.

We adopt the units of $\hbar=k_{\rm B}=1$ throughout.

\section{Expressions of Thermodynamic Potentials}

On the basis of the LW functional for the grand thermodynamic potential,\cite{LW60}
we obtain expressions of basic thermodynamic potentials in a form suitable for numerical studies. 
The normal and superconducting phases are considered separately to present a clear derivation.

\subsection{Normal state}

\subsubsection{Grand thermodynamic potential}
The LW functional for the grand thermodynamic potential $\Omega$ is given for the normal state by
\begin{align}
\Omega = -T\,{\rm Tr}\bigl[\ln \bigl(-\underline{G}_0^{-1}+\underline{\Sigma}\bigr)+\underline{\Sigma}\,\underline{G}\bigr]+\Phi,
\label{LW}
\end{align}
where $T$ is the temperature, ${\rm Tr}$ denotes trace, $\underline{G}_0^{-1}$ is inverse of the non-interacting Green's function matrix, $\underline{\Sigma}$ is the self-energy matrix, 
and $\Phi=\Phi[\underline{G}]$ consists of skeleton-diagram contributions to $\Omega$ in the perturbative treatment of the 
interaction with $\underline{G}_0$ replaced by full Green's function $\underline{G}$.\cite{LW60}
The argument of $\underline{G}$ is given in the coordinate representation as
 $\bigl(\underline{G}\bigr)_{11'}=G(1,1')$ with $1\equiv ({\bf r}_1,\sigma_1,\tau_1)$,\cite{KB62}
 where ${\bf r}_1$ specifies the coordinate, $\sigma_1\equiv \pm \frac{1}{2}$ is the spin quantum number, and 
 $\tau_1$ denotes the imaginary time that runs over $0\leq \tau_1\leq T^{-1}$.
 The self-energy $\underline{\Sigma}$ in Eq.\ (\ref{LW}) is considered as a functional of $\underline{G}$,
 whereas $\underline{G}_0^{-1}$ is treated as given and independent of $\underline{G}$.

The evaluation of Tr in Eq.\ (\ref{LW}) with respect to $\tau$ needs to be performed with a special care,
especially for the logarithmic term.
It was shown based on the path-integral formulation of quantum field theory of many-particle systems\cite{NO88}
that Eq.\ (\ref{LW}) for noninteracting systems reproduces the well-known expression
\begin{align}
\Omega_{0} \equiv &\,-T\,{\rm Tr}\,\ln \bigl(-\underline{G}_{0}^{-1}\bigr)
\notag \\
=&\, -T\, \sum_{{\bf k}\sigma}\ln \bigl(1+e^{-\xi_{\bf k}/T}\bigr),
\label{LW0}
\end{align}
given in terms of the eigenvalue $\xi_{\bf k}$ of the corresponding 
noninteracting Hamiltonian measured from the chemical potential $\mu$.
Using Eq.\ (\ref{LW0}), we express Eq.\ (\ref{LW}) as 
\begin{align}
\Omega=(\Omega-\Omega_{0})+\Omega_{0},
\label{Omega-decomp}
\end{align}
so that $\Omega-\Omega_{0}$ can be expanded safely in the Matsubara frequencies
$\varepsilon_{n}\equiv (2n+1)\pi T$ ($n\equiv 0,\pm 1,\cdots$).
We also focus on the homogeneous attractive Hubbard model with no spin polarization investigated previously,\cite{Kita24,Kita25} 
for which every matrix in Eq.\ (\ref{LW}) becomes diagonal in the Bloch bases.
The corresponding $\Omega$ can be written by using Eqs.\ (\ref{LW0}) and (\ref{Omega-decomp}) as
\begin{align}
\Omega =&\, -2T\sum_{\vec{k}}\left[\ln \frac{-G_0^{-1}(\vec{k})+\Sigma(\vec{k})}{-G_{0}^{-1}(\vec{k})}+\Sigma(\vec{k})G(\vec{k})\right]e^{i\varepsilon_n 0_+}
\notag \\
&\, +\Phi +\Omega_{0},
\label{Omega-2}
\end{align}
where $G_0^{-1}(\vec{k})\equiv i\varepsilon_n-\xi_{\bf k}$ with ${\bf k}$ the Bloch wave vector and $\vec{k}\equiv ({\bf k},i\varepsilon_n)$, $0_+$ is an infinitesimal positive constant, and the factor $2$ originates from the spin degrees of freedom.
The necessity of the factor $e^{i\varepsilon_n 0_+}$ in the evaluation of $\Sigma(\vec{k})G(\vec{k})$
is well-known,\cite{LW60} but it is also required for the logarithmic term even at this stage.
This may be realized from the fact that the logarithm approaches zero slowly in proportion to $(i\varepsilon_n)^{-1}$
for $|\varepsilon_n|\rightarrow\infty$
because of the Hartree-Fock contribution to the self-energy that does not depend on $\varepsilon_n$.

It is desirable to remove the factor $e^{i\varepsilon_n 0_+}$ with numerical ambiguity from Eq.\ (\ref{Omega-2}).
To this end, we express the self-energy as
\begin{align}
\Sigma(\vec{k})=\Sigma_1+\Sigma_{\rm c}(\vec{k}),
\label{Sigma-decomp}
\end{align}
where $\Sigma_1$ is the Hartree-Fock (or first-order) self-energy that is independent of $\varepsilon_n$ generally and also of ${\bf k}$ for the Hubbard model,
and $\Sigma_{\rm c}(\vec{k})$ is the correlation part that vanishes for $|\varepsilon_n|\rightarrow\infty$ 
as shown by using its Lehmann representation.\cite{FW71}
Using the decomposition, we can transform the logarithmic contribution in Eq.\ (\ref{Omega-2}) as
\begin{align}
\Omega_{\ln}\equiv&\, -2T\,\sum_{\vec{k}}e^{i\varepsilon_n 0_+}\ln \frac{-G_0^{-1}(\vec{k})+\Sigma(\vec{k})}{-G_{0}^{-1}(\vec{k})}
\notag \\
=&\, -2T\,\sum_{\vec{k}}e^{i\varepsilon_n 0_+}\left[\ln \frac{-G_0^{-1}(\vec{k})+\Sigma(\vec{k})}{-G_0^{-1}(\vec{k})+\Sigma_1}
\right.
\notag \\
&\,
\hspace{23.5mm}
\left.+ \ln \frac{-G_0^{-1}(\vec{k})+\Sigma_1}{-G_{0}^{-1}(\vec{k})}\right]
\notag \\
=&\, -2T\,\sum_{\vec{k}}\ln \frac{-G_0^{-1}(\vec{k})+\Sigma(\vec{k})}{-G_0^{-1}(\vec{k})+\Sigma_1}
\notag \\
&\,
-2T\sum_{{\bf k}}\ln \frac{1+e^{-(\xi_{\bf k}+\Sigma_1)/T}}{1+e^{-\xi_{\bf k}/T}},
\label{Omega_ln}
\end{align}
where we have used Eq.\ (\ref{LW0})  and also omitted the factor $e^{i\varepsilon_n 0_+}$ 
from the first term on the right-hand side; this logarithm now behaves as $(i\varepsilon_n)^{-2}$ for $|\varepsilon_n| \rightarrow\infty$ so that the factor $e^{i\varepsilon_n 0_+}$ does not affect the result at all.

Using Eq.\ (\ref{Sigma-decomp}), we can also transform the contribution of $\Sigma(\vec{k})G(\vec{k})$ in Eq.\ (\ref{Omega-2}) as
\begin{align}
\Omega_{\Sigma}\equiv &\, -2T\sum_{\vec{k}}\Sigma(\vec{k})G(\vec{k})\,{\rm e}^{i\varepsilon_n 0_+}
\notag \\
=&\,-\Sigma_1 {\cal N}_{\rm e}-2T\sum_{\vec{k}}\Sigma_{\rm c}(\vec{k})G(\vec{k}),
\label{Omega_Sigma}
\end{align}
where
\begin{align}
{\cal N}_{\rm e}=2T\sum_{\vec{k}}G(\vec{k}) \,e^{i\varepsilon_n0_+}
\label{calN_e}
\end{align}
denotes the electron number. 
Finally, we write $\Phi$ as a sum of the Hartree-Fock and  correlation contributions as
\begin{align}
\Phi=\Phi_1+\Phi_{\rm c}=\frac{1}{2}\Sigma_1{\cal N}_{\rm e} +\Phi_{\rm c}.
\label{Phi-decompose}
\end{align}
Substituting Eqs.\ (\ref{Omega_ln}), (\ref{Omega_Sigma}), and (\ref{Phi-decompose}) in Eq.\ (\ref{Omega-2}), we obtain
\begin{align}
\Omega= &\, -2T\,\sum_{\vec{k}}\left[\ln \frac{-G_0^{-1}(\vec{k})+\Sigma(\vec{k})}{-G_0^{-1}(\vec{k})+\Sigma_1}
+\Sigma_{\rm c}(\vec{k})G(\vec{k})\right]+\Phi_{\rm c}
\notag \\
&\, -2T\sum_{{\bf k}}\ln\Bigl[1+e^{-(\xi_{\bf k}+\Sigma_1)/T}\Bigr]-\frac{1}{2}\Sigma_1{\cal N}_{\rm e} .
\label{Omega-Num}
\end{align}
This expression could have been obtained directly from Eq.\ (\ref{Omega-2}) by adding and subtracting the grand thermodynamic potential $\Omega_1$ of the Hartree-Fock approximation.
However, the rather redundant derivation given above may also have the advantage of clarifying the subtlety
required for evaluating the logarithmic term with the factor $e^{i\varepsilon_n0_+}$.

The functional derivative $\delta\Omega/\delta G(\vec{k})$ of Eq.\ (\ref{Omega-2}) can be calculated as
\begin{align}
\frac{\delta\Omega}{\delta G(\vec{k})}=&\,2T\sum_{\vec{k}'}\Bigl\{\Bigl[G_0^{-1}(\vec{k}')-\Sigma(\vec{k}')\Bigr]^{-1}-G(\vec{k}')\Bigr\}
\notag \\
&\,\times \frac{\delta \Sigma(\vec{k}')}{\delta G(\vec{k})} e^{i\varepsilon_{n'}0_+}
-2T \Sigma(\vec{k})+\frac{\delta\Phi}{\delta G(\vec{k})}.
\end{align}
Hence, the stationarity condition
$\delta \Omega/\delta G(\vec{k})=0$ on Eq.\ (\ref{Omega-2}),
which results naturally from the Legendre transformation procedure of $\Omega$,\cite{DDM64,Kita14} 
yields Dyson's equation
\begin{align}
G(\vec{k})=\Bigl[G_0^{-1}(\vec{k})-\Sigma(\vec{k})\Bigr]^{-1},
\label{Dyson}
\end{align}
with the self-energy given by
\begin{align}
\Sigma(\vec{k})=&\,\frac{1}{2T}\frac{\delta\Phi}{\delta G(\vec{k})}.
\label{Sigma-Phi_n}
\end{align}
We can calculate $\Omega$ by Eq.\ (\ref{Omega-Num}) numerically
based on the self-consistent solution of Eqs.\ (\ref{Dyson}) and (\ref{Sigma-Phi_n}) 
obtained under the condition (\ref{calN_e}) for a given $\Phi$.

\subsubsection{$\Phi$ functional}

As in the previous studies,\cite{Kita24,Kita25} we adopt the fluctuation-exchange (FLEX) approximation\cite{BSW89,BS89} 
for $\Phi$ in the normal state given by
\begin{align}
\Phi=&\,\Phi_1-\frac{1}{2} T\sum_{\vec{q}}\bigl[U\chi_{GG}^{}(\vec{q}\,)\bigr]^2
\notag \\
&\, +T\sum_{\vec{q}}\Bigl[\ln(1-x)+x+\frac{1}{2}x^2\Bigr]_{x=U\chi_{G\bar{G}}^{}(\vec{q}\,)} 
\notag \\
&\,  +\frac{1}{2}T\sum_{\vec{q}}\biggl[\ln(1+x)-x+\frac{1}{2}x^2\biggr]_{x=U\chi_{GG}^{}(\vec{q}\,)}
\notag \\
&\,  +\frac{3}{2}T\sum_{\vec{q}}\biggl[\ln(1-x)+x+\frac{1}{2}x^2\biggr]_{x=U\chi_{GG}^{}(\vec{q}\,)}\, .
\label{Phi_n}
\end{align}
Here $\Phi_1$ is the Hartree-Fock contribution in Eq.\ (\ref{Phi-decompose}), and $\chi_{GG}^{}(\vec{q}\,)$ and $\chi_{G\bar{G}}^{}(\vec{q}\,)$
are defined in terms of $G(\vec{k})$ and its time-reversed partner
$\bar{G}(\vec{k})\!\equiv\! G(-\vec{k})$
generally by
\begin{align}
\chi_{AB}^{}(\vec{q}\,)\equiv&\, -\frac{T}{{\cal N}_{\rm a}}\sum_{\vec{k}}A(\vec{k}+\vec{q}\,)B(\vec{k}) ,
\label{chi_AB-def}
\end{align}
with $\vec{q}\equiv({\bf q},i\omega_\ell)$ denoting the four-momentum consisting of the wave vector ${\bf q}$ 
and boson Matsubara frequency $\omega_\ell$.

The $\Phi$ functional above should be supplemented in the pseudogap phase by the additional contribution\cite{Kita24,Kita25}
\begin{align}
\varDelta\Phi_\lambda \equiv \lambda {\cal N}_{\rm e}\bigl[1-U\chi_{G\bar{G}}^{}(\vec{0})\bigr],
\label{Phi_lambda^n}
\end{align}
to keep $U\chi_{G\bar{G}}(\vec{0})=1$.
Though it affects the form of the self-energy, this $\varDelta\Phi_\lambda$ vanishes in the calculation of 
Eq.\ (\ref{Omega-Num}) based on the solution of Dyson's equation because of $U\chi_{G\bar{G}}(\vec{0})=1$.

It should also be noted that Eq.\ (\ref{Phi_n}) has been slightly modified in form from Eq.\ (3) of Ref.\ \onlinecite{Kita24}
by using the identity
\begin{align}
-\frac{T}{2}\sum_{\vec{q}}\bigl[U\chi_{GG}(\vec{q})\bigr]^2=-\frac{T}{2}\sum_{\vec{q}}\bigl[U\chi_{G\bar{G}}(\vec{q})\bigr]^2,
\label{Phi_2}
\end{align}
which holds for the second-order contribution to $\Phi$ in the normal state.
However, the former expression with $\chi_{GG}(\vec{q})$ is far more suitable numerically because of the inequality
$|\chi_{GG}(\vec{q})|\ll |\chi_{G\bar{G}}(\vec{q})|$ that holds for the attractive Hubbard model
at low temperatures.
Indeed, Eq.\ (\ref{Phi_2}) indicates that the tiny second-order contribution given in terms of the particle-hole bubble $\chi_{GG}(\vec{q})$
results by the expression in terms of the particle-particle bubble $\chi_{G\bar{G}}(\vec{q})$ through the cancelation of its huge real and imaginary numbers.
Hence, we have calculated the normal-state $\Phi$ by Eq.\ (\ref{Phi_n}) above.

\subsubsection{Other thermodynamic potentials\label{subsubsec:OTP-n}}

The Helmholtz free energy ${\cal F}$ is obtained from Eq.\ (\ref{Omega-Num}) by the thermodynamic relation
\begin{align}
{\cal F}=\Omega+\mu{\cal N}_{\rm e},
\label{calF}
\end{align}
where ${\cal N}_{\rm e}$ is the electron number defined by Eq.\ (\ref{calN_e}).
Second, the internal energy ${\cal E}$ can be calculated by the expression 
\begin{subequations}
\label{calE}
\begin{align}
{\cal E}=2T\sum_{\vec{k}}\left[(\xi_{\bf k}+\mu)G(\vec{k}) +\frac{1}{2}\Sigma(\vec{k})G(\vec{k})\right]
e^{i\varepsilon_n 0_+},
\label{calE1}
\end{align}
resulting from thermodynamic average of the Hamiltonian,\cite{LW60}
where the first and second terms in the square brackets represent the kinetic and interaction energies, respectively;
the latter can be transformed in the same way as Eq.\ (\ref{Omega_Sigma}).
Alternatively, we may use the thermodynamic relation
\begin{align}
{\cal E}=-T^2\frac{\partial({\cal F}/T)}{\partial T},
\label{calE2}
\end{align}
\end{subequations}
to obtain ${\cal E}$ from Eq.\ (\ref{calF}) by numerical differentiation.
Third, entropy ${\cal S}$ can be calculated by either of the thermodynamic relations
\begin{subequations}
\label{calS}
\begin{align}
{\cal S}=&\,\frac{{\cal E}-{\cal F}}{T}
\label{calS1}
\\
=&\, -\frac{\partial {\cal F}}{\partial T} .
\label{calS2}
\end{align}
\end{subequations}
Finally, the heat capacity ${\cal C}_V$ can be obtained by
\begin{subequations}
\label{calC}
\begin{align}
{\cal C}_V=&\,\frac{\partial {\cal E}}{\partial T} 
\label{calC1}
\\
=&\, T\frac{\partial {\cal S}}{\partial T} .
\label{calC2}
\end{align}
\end{subequations}
Existence of two apparently different expressions in each of Eqs.\ (\ref{calE})-(\ref{calC})
enables us to check the accuracy of the obtained numerical results.

\subsection{Superconducting phase}
\subsubsection{Grand thermodynamic potential}

For superconducting phases with spin-singlet pairing, Eq.\ (\ref{Omega-2}) is replaced by\cite{Kita11,Kita96}
\begin{align}
\Omega =&\, -T\sum_{\vec{k}}{\rm Tr}\,\biggl\{\ln \Bigl[-\hat{G}_0^{-1}(\vec{k})+\hat\Sigma(\vec{k})\Bigr]
-\ln\Bigl[-\hat{G}_{0}^{-1}(\vec{k})\Bigr]
\notag \\
&\, +\hat\Sigma(\vec{k})\hat{G}(\vec{k})\biggr\}\,\hat{1}(\varepsilon_n) 
+\Phi +\Omega_{0},
\label{Omega_s}
\end{align}
where $\hat{G}_0^{-1}$, $\hat\Sigma$, $\hat{G}$, and $\hat{1}(\varepsilon_n)$ are Nambu matrices defined by
\begin{subequations}
\label{hatG_0^-1-hatG}
\begin{align}
\hat{G}_0^{-1}(\vec{k})=\begin{bmatrix} i\varepsilon_n-\xi_{\bf k} & 0 \\ 0 & 
i\varepsilon_n+\xi_{\bf k}\end{bmatrix},
\label{hatG_0^-1}
\end{align}
\begin{align}
\hat\Sigma(\vec{k})=\begin{bmatrix} \Sigma(\vec{k}) & \Delta(\vec{k}) \\ \Delta(\vec{k}) & 
-\Sigma(-\vec{k})\end{bmatrix},
\label{hatSigma}
\end{align}
\begin{align}
\hat{G}(\vec{k})=\begin{bmatrix} G(\vec{k}) & F(\vec{k}) \\ F(\vec{k}) & -G(-\vec{k}) \end{bmatrix},
\label{hatG}
\end{align}
\begin{align}
\hat{1}(\varepsilon_n)= \begin{bmatrix}e^{i \varepsilon_n 0_+} & 0 \\ 0 & e^{-i \varepsilon_n 0_+}\end{bmatrix} ,
\end{align}
\end{subequations}
with $\Delta(-\vec{k})=\Delta(\vec{k})$ and $F(-\vec{k})=F(\vec{k})$.

The stationarity conditions
$\delta \Omega/\delta G(\vec{k})=\delta \Omega/\delta F(\vec{k})=0$ on Eq.\ (\ref{Omega_s}) yield the Dyson-Gor'kov equation
\begin{align}
\hat{G}(\vec{k})=\Bigl[\hat{G}_0^{-1}(\vec{k})-\hat\Sigma(\vec{k})\Bigr]^{-1},
\label{DG}
\end{align}
with the self-energies
\begin{subequations}
\label{Sigma-Phi}
\begin{align}
\Sigma(\vec{k})=&\,\frac{1}{2T}\frac{\delta\Phi}{\delta G(\vec{k})},
\label{Sigma-Phi1}
\\
\Delta(\vec{k})=&\,\frac{1}{2T}\frac{\delta\Phi}{\delta F(\vec{k})}.
\label{Sigma-Phi2}
\end{align}
\end{subequations}
Equations (\ref{DG}) and (\ref{Sigma-Phi}) can be derived from Eq.\ (\ref{Omega_s})
similarly as Eqs.\ (\ref{Dyson}) and (\ref{Sigma-Phi_n}) from Eq.\ (\ref{Omega-2}).

Following the transformation of Eq.\ (\ref{Omega-2}) into Eq.\ (\ref{Omega-Num}) for the normal state,
we add and subtract the logarithmic mean-field contribution
\begin{align}
\Omega_{1,\ln}=&\,\Omega_0
-T\sum_{\vec{k}}{\rm Tr}\,\biggl\{\ln \Bigl[-\hat{G}_0^{-1}(\vec{k})+\hat\Sigma_1\Bigr]
\notag \\
&\,-\ln\Bigl[-\hat{G}_{0}^{-1}(\vec{k})\Bigr]\biggr\}\,\hat{1}(\varepsilon_n),
\label{Omega_MF}
\end{align}
where $\hat\Sigma_1$ is defined by
\begin{align}
\hat\Sigma_1\equiv \lim_{|\varepsilon_n|\rightarrow\infty}
\hat\Sigma(\vec{k})\equiv \begin{bmatrix}\Sigma_1 &\Delta_1 \\ \Delta_1 & -\Sigma_1\end{bmatrix} .
\label{hatSigma_1}
\end{align}
Its upper elements can be expressed for the attractive Hubbard model studied previously\cite{Kita25} as
\begin{subequations}
\label{hatSigma_1-def}
\begin{align}
\Sigma_1 = &\,\frac{U}{{\cal N}_{\rm a}}T\sum_{\vec{k}}G(\vec{k}) \,e^{i\varepsilon_n 0_+}=\frac{1}{2}U \frac{{\cal N}_{\rm e}}{{\cal N}_{\rm a}},
\\
\Delta_1=&\, \frac{U}{{\cal N}_{\rm a}}T\sum_{\vec{k}}F(\vec{k}) ,
\label{Delta_1-def}
\end{align}
\end{subequations}
with $U<0$ the interaction strength, ${\cal N}_{\rm a}$ the number of lattice sites, and ${\cal N}_{\rm e}$ is defined by Eq.\ (\ref{calN_e}).
Equation (\ref{hatSigma_1}) is the first-order self-energy given in terms of the fully self-consistent
$\hat{G}$.
It is shown in Appendix A that Eq.\ (\ref{Omega_MF}) can be written alternatively as
\begin{align}
\Omega_{1,\ln}=2\sum_{\bf k}\left[-T\ln\bigl(1+e^{-E_{{\bf k}1}/T}\bigr)+\frac{\xi_{{\bf k}1} -E_{{\bf k}1}}{2}\right] ,
\label{Omega_MF2}
\tag{\ref{Omega_MF}$^\prime$}
\end{align}
with 
\begin{subequations}
\begin{align}
\xi_{{\bf k}1}\equiv&\, \xi_{{\bf k}}+\Sigma_1,
\\
E_{{\bf k}1}\equiv &\,\sqrt{\xi_{{\bf k}1}^2+\Delta_1^2}.
\end{align}
\end{subequations}
We also express
\begin{align}
\hat\Sigma(\vec{k})=\hat\Sigma_1+\hat\Sigma_{\rm c}(\vec{k}),
\label{hSigma-decomp}
\end{align}
and transform the $\hat\Sigma(\vec{k})\hat{G}(\vec{k})$ contribution of Eq.\ (\ref{Omega_s}) into
\begin{align}
\Omega_{\Sigma}\equiv &\,-T\sum_{\vec{k}}{\rm Tr}\,\hat\Sigma(\vec{k})\hat{G}(\vec{k})\hat{1}(\varepsilon_n)
\notag \\
=&\,-\Sigma_1{\cal N}_{\rm e}-2\frac{\Delta_1^2}{U}{\cal N}_{\rm a}-T\sum_{\vec{k}}{\rm Tr}\,\hat\Sigma_{\rm c}(\vec{k})\hat{G}(\vec{k}),
\label{Omega_Sig2}
\end{align}
where we have used Eq.\ (\ref{hatSigma_1-def}).
Finally, we decompose $\Phi$ into the first-order mean-field contribution and correlation one
as
\begin{align}
\Phi = &\,\Phi_1+\Phi_{\rm c}
=\frac{1}{2}\Sigma_1{\cal N}_{\rm e}+
\frac{\Delta_1^2}{U}{\cal N}_{\rm a}+\Phi_{\rm c}.
\label{Phi-decomp_s}
\end{align}
Using Eq.\ (\ref{Omega_MF}), (\ref{Omega_MF2}), (\ref{Omega_Sig2}), and (\ref{Phi-decomp_s}),
we can express Eq.\ (\ref{Omega_s}) in a form suitable for numerical studies as
\begin{align}
\Omega =&\,\Omega_1 -T\sum_{\vec{k}}{\rm Tr}\,\biggl\{\ln \Bigl[-\hat{G}_0^{-1}(\vec{k})+\hat\Sigma(\vec{k})\Bigr]
\notag \\
&\, -\ln\Bigl[-\hat{G}_{0}^{-1}(\vec{k})+\hat\Sigma_1\Bigr]
+\hat\Sigma_{\rm c}(\vec{k})\hat{G}(\vec{k})\biggr\}+\Phi_{\rm c} ,
\label{Omega_s-Num}
\end{align}
where $\Omega_1$ is given by
\begin{align}
\Omega_1=&\,2\sum_{\bf k}\biggl[-T\ln\bigl(1+e^{-E_{{\bf k}1}/T}\bigr)+\frac{\xi_{{\bf k}1}-E_{{\bf k}1}}{2}\biggr]
\notag \\
&\, -\frac{1}{2}\Sigma_1{\cal N}_{\rm e}-\frac{\Delta_1^2}{U}{\cal N}_{\rm a} .
\end{align}

\subsubsection{$\Phi$ functional}

As in the previous studies,\cite{Kita24,Kita25} 
we here adopt the fluctuation-exchange approximation for superconductivity (FLEX-S approximation).\cite{Kita11}
Specifically, the $\Phi$ functional in Eq.\ (\ref{Omega_s}) is given by Eq.\ (12) of Ref.\ \onlinecite{Kita24}, i.e.,
\begin{align}
\Phi 
 =&\, \Phi_1-\frac{1}{2} T\sum_{\vec{q}}\bigl[U\chi_+^{}(\vec{q}\,)\bigr]^2
\notag \\
&\, +\frac{1}{2}T\sum_{\vec{q}}{\rm Tr}\,\biggl[\ln\bigl(\underline{1}+\underline{x}\bigr)-\underline{x}+\frac{1}{2}\underline{x}^2\biggr]_{\underline{x}=U\underline{\chi}^{({\rm c})}(\vec{q}\,)}
\notag \\
&\, +\frac{3}{2}T\sum_{\vec{q}}\biggl[\ln\bigl(1-x\bigr)+x+\frac{1}{2}x^2\biggr]_{x=U\chi_+^{}(\vec{q}\,)}.
\label{Phi}
\end{align}
Here $\Phi_1$ is the mean-field contribution in Eq.\ (\ref{Phi-decomp_s}), $\underline{1}$ is the $3\times 3$ unit matrix, $\underline{\chi}^{({\rm c})}(\vec{q}\,)$ is
defined in terms of Eq.\ (\ref{chi_AB-def}) with $A,B\!=\! G,\bar{G},F$ by 
\begin{subequations}
\begin{align}
\underline{\chi}^{({\rm c})}(\vec{q}\,)\equiv 
\begin{bmatrix}
\vspace{1mm}
\chi_{-}^{}(\vec{q}\,)  & \sqrt{2}\chi_{GF}^{}(\vec{q}\,) & -\sqrt{2}\chi_{\bar{G}F}^{}(\vec{q}\,)\\
\vspace{1mm}
\sqrt{2}\chi_{GF}^{}(\vec{q}\,) & -\chi_{G\bar{G}}^{}(\vec{q}\,) & -\chi_{FF}^{}(\vec{q}\,) \\
-\sqrt{2}\chi_{\bar{G}F}^{}(\vec{q}\,) & -\chi_{FF}^{}(\vec{q}\,) & -\chi_{\bar{G}G}^{}(\vec{q}\,) 
\end{bmatrix},
\label{chi^(0c)}
\end{align}
and $\chi_\pm^{}(\vec{q}\,)$ denote
\begin{align}
\chi_\pm^{}(\vec{q}\,)\equiv &\,\chi_{GG}^{}(\vec{q}\,)\pm \chi_{FF}^{}(\vec{q}\,) .
\end{align}
\end{subequations}
Equation (\ref{Phi}) for $F\rightarrow 0$ reduces to Eq.\ (\ref{Phi_n}), as it should.

The functional $\Phi$ above should be supplemented by adding
\begin{align}
\varDelta\Phi_\lambda=\lambda {\cal N}_{\rm e}\left[1-U\chi_{G\bar{G}}(\vec{0})-U\chi_{FF}(\vec{0})\right] ,
\label{Phi_lambda}
\end{align}
before performing the differentiations of Eq.\ (\ref{Sigma-Phi}).\cite{Kita25} However, this contribution vanishes
in the evaluation of Eq.\ (\ref{Omega_s}) owing to $U\chi_{G\bar{G}}(\vec{0})+U\chi_{FF}(\vec{0})=1$.

\subsubsection{Other thermodynamic potentials}

Equations (\ref{calF})-(\ref{calC}) for the normal state 
remain valid in the superconducting phase except Eq.\ (\ref{calE1}) for the internal energy ${\cal E}$,
for which the anomalous contribution should be added as
\begin{align}
{\cal E}=&\,2T\sum_{\vec{k}}\biggl\{\left[(\xi_{\bf k}+\mu)G(\vec{k}) +\frac{1}{2}\Sigma(\vec{k})G(\vec{k}) \right]
e^{i\varepsilon_n 0_+}
\notag \\
&\,
\hspace{10mm}+\frac{1}{2}\Delta(\vec{k})F(\vec{k})\Biggr\} .
\label{calE1-s}
\tag{\ref{calE1}$^\prime$}
\end{align}
This expression can be proved based on the equations of motion for the field operators
in the same way as that for condensed Bose systems.\cite{Kita09}

\section{Numerical Results}

\begin{figure}[b]
\centering
\includegraphics[width=0.9\linewidth]{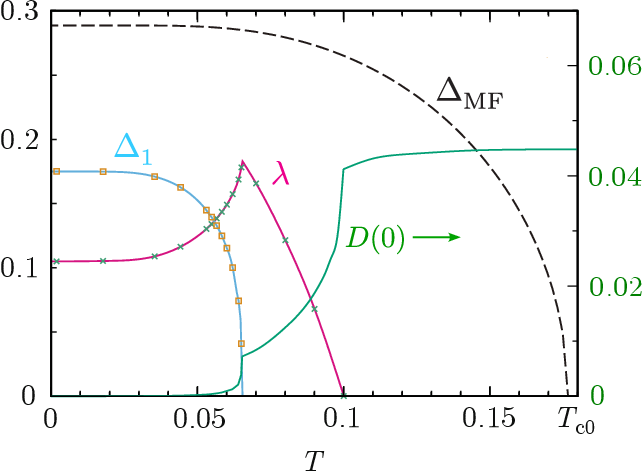}
\caption{\label{fig1}Temperature dependences of the Lagrange multiplier $\lambda$,
the first-order anomalous self-energy $\Delta_1^{}$ in comparison with the mean-field energy gap $\Delta_{\rm MF}^{}$, and the single-particle density of states at zero energy $D(0)$.
}
\end{figure}

We have performed a fully self-consistent numerical calculation of the thermodynamic potentials
for the low-density  attractive Hubbard model considered previously.\cite{Kita24,Kita25}
Specifically, the band structure is given by $\xi_{\bm k}=k^2-\mu$ with $0\leq k\leq 10$ in units of $2m\!=\!k_{\rm F}\!=\!\varepsilon_{\rm F}^0=1$, 
where $m$ is the electron mass, $k_{\rm F}$ is the Fermi momentum, and $\varepsilon_{\rm F}^0=\mu_0(T=0)$ denotes the non-interacting Fermi energy.
Temperature dependences of basic parameters were obtained as Fig.\ \ref{fig1}
at an intermediate coupling of $UN(0)=-0.09$,\cite{Kita24} where $\Delta_{\rm MF}$ and 
$T_{{\rm c}0}\!=\!0.1767$ are the mean-field results on the energy gap and transition temperature,
$\lambda$ is the prefactor of the 1PR contribution to the self-energy in Eq.\ (29) of Ref.\ \onlinecite{Kita25},
$\Delta_1^{}$ is the energy gap evaluated by Eq.\ (\ref{Delta_1-def}) in terms of the fully self-consistent $F$, and $D(0)$ denotes the single-particle density of states at zero energy defined by Eq.\ (40) of Ref.\ \onlinecite{Kita25}.
The parameters $\lambda$ and $\Delta_1$ start to develop at $T_{\rm 1PR}=0.100$ and $T_{\rm c}=0.0654$, respectively.
Whereas $T_{\rm c}$ signifies the superconducting transition temperature, 
$T_{\rm 1PR}$ is not directly connected with any broken symmetry but only characterized
by the emergence of the 1PR structure in the self-energy.\cite{Kita24}
As shown in a previous study,\cite{Kita25} 
this $T_{\rm 1PR}$ marks the onset of a pseudogap behavior in the single-particle density of states,
as seen in a sharp drop of $D(0)$ at $T_{\rm 1PR}$ in Fig.\ \ref{fig1};
it may be denoted more appropriately as $T_{\rm pg}$ based on the observable characteristic.
Given below are temperature dependences of the basic thermodynamic potentials
corresponding to the phase diagram of Fig.\ \ref{fig1}.

\begin{figure}[t]
\centering
\includegraphics[width=0.92\linewidth]{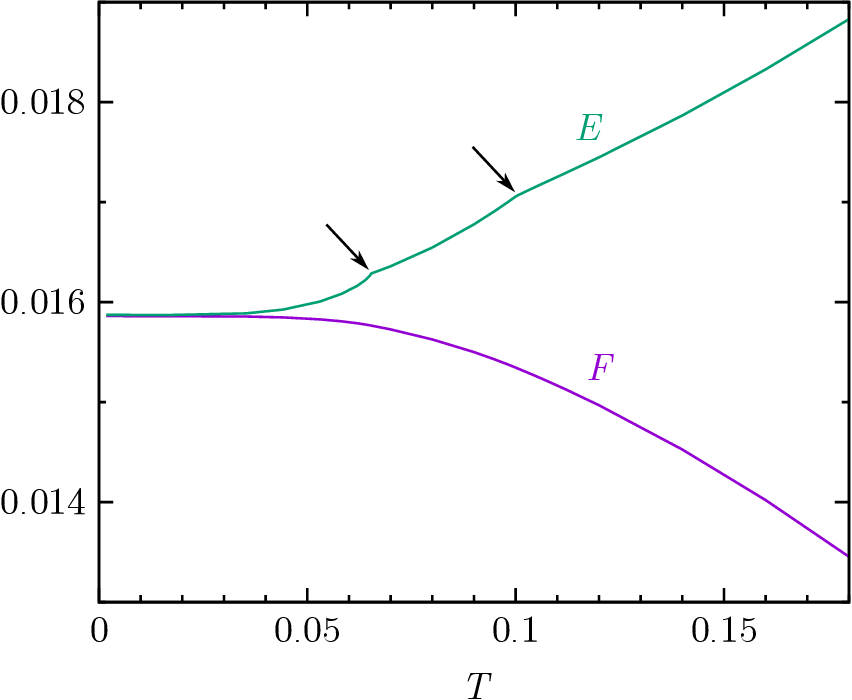}
\caption{\label{fig2}Helmholtz free energy $F$ and internal energy $E$ per unit volume
defined by Eq.\ (\ref{FE-def}) as a function of temperature $T$.
The arrows indicate cusps in the curve of $E$.
}
\end{figure}
\begin{figure}[t]
\centering
\hspace{5mm}\includegraphics[width=0.9\linewidth]{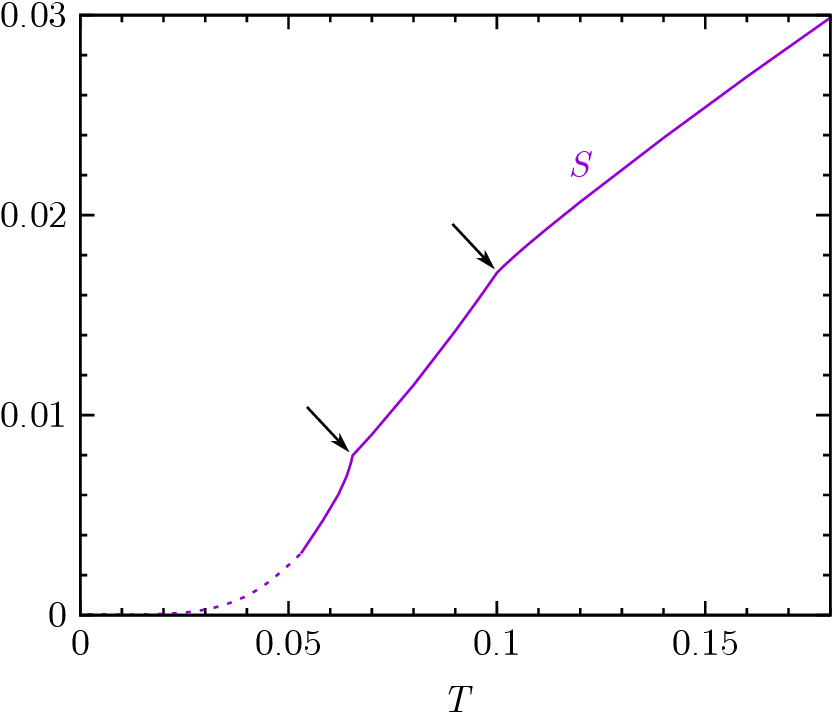}
\caption{\label{fig3}Entropy per unit volume as a function of temperature.
The full and dotted lines were calculated by Eqs.\ (\ref{calS1}) and (\ref{calS2}), respectively.
The arrows indicate cusps in the curve.}
\end{figure}

\begin{figure}[t]
\centering
\includegraphics[width=0.95\linewidth]{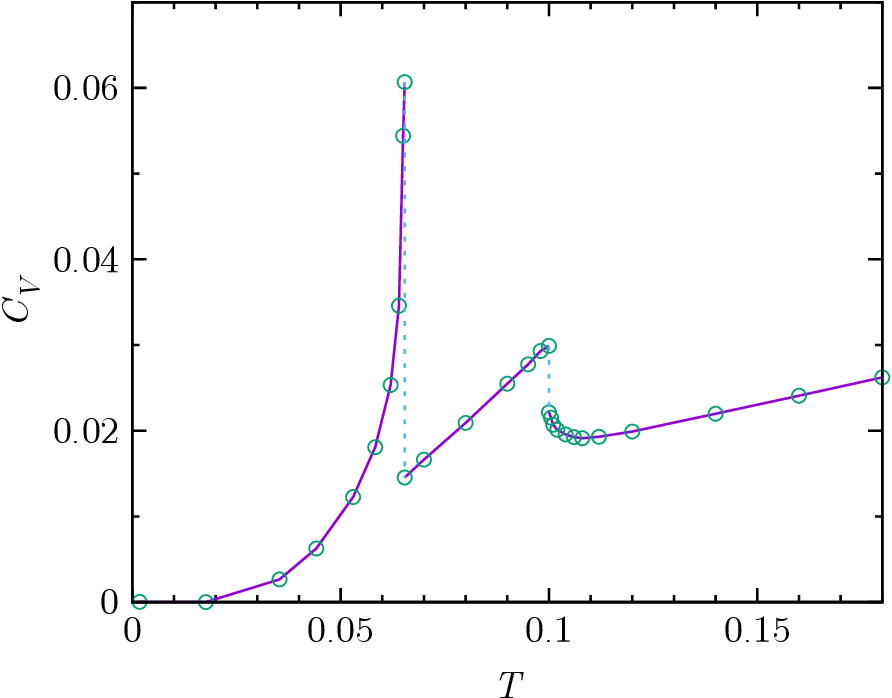}
\caption{\label{fig4}Temperature dependence of the specific heat $C_V$ per unit volume
calculated by Eq.\ (\ref{calC1}). The circles are data points, and lines are obtained by linearly interpolating
every adjacent data points, and the broken lines are the interpolations of two discontinuities at
$T=T_{\rm c}$ and $T_{1{\rm PR}}$.
}
\end{figure}

Figure \ref{fig2} plots the Helmholtz free energy $F$ and internal energy $E$ per unit volume
as a function of temperature $T$, where the Hartree-Fock energy giving a constant shift has been subtracted.
Specifically, $F$ and $E$ in Fig.\ \ref{fig2} are defined by 
\begin{align}
F \equiv \frac{{\cal F}-U{\cal N}_{\rm e}^2/4{\cal N}_{\rm a}}{{\cal V}}, \hspace{5mm} 
E \equiv \frac{{\cal E}-U{\cal N}_{\rm e}^2/4{\cal N}_{\rm a}}{{\cal V}},
\label{FE-def}
\end{align}
where ${\cal V}$ denotes the volume given in terms of the volume of the unit cell  $V_{\rm unit}$ by
${\cal V}={\cal N}_{\rm a}V_{\rm unit}$. This potential $F$ has been calculated by Eq.\ (\ref{calF})
in terms of $\Omega$ given by Eqs.\ (\ref{Omega-Num}) and (\ref{Omega_s-Num}) for the normal and superfluid phases, respectively,
whereas $E$ has been obtained based on Eqs.\ (\ref{calE1}) and 
(\ref{calE1-s}) for the normal and superfluid phases, respectively.
We observe that $F$ and $E$ approach a common value $\approx\! 0.0159$
for $T\rightarrow 0$ to within an error of $0.1\%$,
in agreement with the third law of thermodynamics.
The reduction of this value $0.0159$ from the non-interacting kinetic energy $(5\pi^2)^{-1}\approx 0.0203$ 
per volume at $T=0$ in the present units is caused by correlations.
Whereas $F$ is smooth over the whole temperature range,
the internal energy $E$ exhibits cusps characteristic of continuous transitions at $T_{\rm c}=0.0654$ and $T_{\rm 1PR}=0.100$.
Thus, the emergence of the 1PR structure in the self-energy at $T=T_{\rm 1PR}$ can be regarded 
as a kind of continuous transitions.

The cusps at $T_{\rm c}=0.0654$ and $T_{\rm 1PR}=0.100$ can be seen more clearly in the entropy curve of Fig.\ \ref{fig3}
obtained by Eq.\ (\ref{calS1}) for $T\gtrsim 0.053$;
the dotted curve at lower temperatures is based on Eq.\ (\ref{calS2}) which is more reliable
numerically at vanishing temperatures.

The cusps in Figs.\ \ref{fig2} and \ref{fig3} have been confirmed to
yield the discontinuities in the specific heat per unit volume as given by Fig.\ \ref{fig4};
the whole curve was obtained by Eq.\ (\ref{calC1}).
The discontinuity at $T=T_{1{\rm PR}}$ is seen to be smaller in magnitude
than that at the superconducting transition $T=T_{{\rm c}}$.
A crude analytic consideration given in Appendix\ref{AppB} clarifies that 
the former is caused by the linear emergence of $\lambda$ seen in Fig.\ \ref{fig1}.
As a reference, 
the corresponding specific heat of the non-interacting gas is given in the present units
by $C_{V0}=T/6$ with $C_{V0}(0.18)=0.6$.
More detailed features of the $C_V$ curve are summarized as follows.
First, we observe an upward turn of $C_V$ upon approaching $T=T_{1{\rm PR}}$ 
from above, which results from the singular behavior of the normal self-energy for 
$U\chi_{G\bar{G}}(\vec{0})\rightarrow 1$.
Second, the curve in the superconducting phase may be diverging towards $T_{{\rm c}}$
according to the behavior of the $\Delta_1$ curve in Fig.\ \ref{fig1};
we cannot give any confirming statement on this issue from the present numerical study,
which needs to be clarified in the future.

Finally, we comment on the accuracy of the present numerical results on the thermodynamic potentials.
The summations over the boson Matsubara frequencies 
of Eqs.\ (\ref{Phi_n}) and (\ref{Phi}) were performed as described in Appendix B of Ref.\ \onlinecite{Kita24}
by increasing the number $N_\omega$ of integration points up to $N_\omega=400$
for obtaining the potentials with enough accuracy.
As stated previously, existence of two different expressions in each of Eqs.\ (\ref{calE})--(\ref{calC})
enables us to estimate the numerical errors in the curves of Figs.\ \ref{fig2}--\ref{fig4}.
The differentiations of the formulas have been performed based on the natural cubic splines\cite{NRC} of the data points 
constructed separately for each of the normal, pseudogap, and superconducting phases.
Comparing the results of Eqs.\ (\ref{calE1}) and (\ref{calE2}) at intermediate points, we estimate the relative errors of
the curves in Fig.\ \ref{fig2} to be of order of $0.1\%$ for $T\gtrsim0.05$. 
Similarly, the relative errors of the entropy curve in Fig.\ \ref{fig3} and specific heat curve in Fig.\ \ref{fig4} are of order of $1\%$ for $T\gtrsim0.05$.

\section{Summary}

We have performed a fully self-consistent numerical study of
thermodynamic potentials for the attractive Hubbard model considered previously.\cite{Kita24,Kita25}
To the best of our knowledge, this is the first time where thermodynamic potentials have been calculated fully
self-consistently and continuously through transitions based on the LW functional with correlations.
It is thereby shown that the pseudogap phase characterized by the emergence of the 1PR structure in the self-energy
exhibits singularities in thermodynamic potentials at the threshold temperature,
specifically, cusps in the internal energy and entropy and discontinuities in the heat capacity,
as seen in Figs.\ \ref{fig2}-\ref{fig4}.
Unlike continuous phase transitions with broken symmetries,
this pseudogap transition accompanies no apparent broken symmetries.
Together with the pseudogap behavior in the single-particle density of states,\cite{Kita25}
this feature can be used for the experimental identification
of the pseudogap phases realized by the present scenario.

\appendix

\section{Equality between Eqs.\ (\ref{Omega_MF}) and (\ref{Omega_MF2})}

To show the equality between Eqs.\ (\ref{Omega_MF}) and (\ref{Omega_MF2}),
we consider the following function
\begin{align}
f(\xi)\equiv &\, -T \, \sum_{n}\biggl\{{\rm Tr}\, \hat{1}(\varepsilon_n)
\ln \begin{bmatrix} - i\varepsilon_n+\xi_1 \!\!&\!\!\Delta_1 \\ \Delta_1 \!\!&\!\! -i\varepsilon_n-\xi_1\end{bmatrix}
\notag \\
&\,-2 e^{i\varepsilon_n 0_+} \ln (-i\varepsilon_n+\xi)\biggr\}
\label{f(xi)-1}
\end{align}
with $\xi_1\equiv\xi+\Sigma_1$,
which satisfies $f(\infty)=0$.
Its derivative can be transformed as
\begin{align}
f'(\xi)=&\, -T \, \sum_{n}\biggl\{{\rm Tr}\, \hat{1}(\varepsilon_n)\!
\begin{bmatrix} - i\varepsilon_n+\xi_1 \!\!&\!\! \Delta_1 \\ \Delta_1 \!\!&\!\! -i\varepsilon_n-\xi_1\end{bmatrix}^{-1}\!\hat\tau_3
\notag \\
&\,-2 \frac{e^{i\varepsilon_n 0_+}}{-i\varepsilon_n+\xi}\biggr\}
\notag \\
=&\, 2T\sum_n \left[\frac{i\varepsilon_n+\xi_1}{(i\varepsilon_n)^2-\xi_1^2-E_1^2}-\frac{1}{i\varepsilon_n-\xi}\right]e^{i\varepsilon_n 0_+}
\notag \\
=&\, 2T\sum_n \left(\frac{u_1^2}{i\varepsilon_n-E_1}+\frac{v_1^2}{i\varepsilon_n+E_1}-\frac{1}{i\varepsilon_n-\xi}\right)e^{i\varepsilon_n 0_+}
\notag \\
=&\, 2\frac{\xi_1}{E_1}\bar{n}(E_1)+1-\frac{\xi_1}{E_1}-2\bar{n}(\xi),
\end{align}
with $E_1\!\equiv\!\sqrt{\xi_1^2+\Delta_1^2}$, $u_1^2\!\equiv\! \frac{1}{2}\left(1+\frac{\xi_1}{E_1}\right)$, $v_1^2\!\equiv \!\frac{1}{2}\left(1-\frac{\xi_1}{E_1}\right)$, and $\bar{n}(E_1)\equiv(e^{E_1/T}+1)^{-1}$.
Integrating $f'(\xi)$ with the condition $f(\infty)=0$,
we obtain
\begin{align}
f(\xi) = -2T\ln\frac{1+e^{-E_1/T}}{1+e^{-\xi/T}}+\xi_1-E_1.
\label{f(xi)-2}
\tag{\ref{f(xi)-1}$^\prime$}
\end{align}
Thus, we have proved the equality between Eqs.\ (\ref{f(xi)-1}) and (\ref{f(xi)-2}), which implies the equality between Eqs.\ (\ref{Omega_MF}) and (\ref{Omega_MF2}).

\section{Some analytic expressions at $T_{1{\rm PR}}$\label{AppB}}

We here derive crude analytic expressions for the initial slope of $\lambda(T)$
and the discontinuity of the specific heat $\varDelta C_V$ at $T= T_{1{\rm PR}}$.

\subsection{Initial slope of $\lambda(T)$}

Let us express Green's function in the pseudogap phase of $\lambda\neq 0$ 
as $G_\lambda(\vec{k})$.
This phase is characterized by the identity\cite{Kita25}
\begin{align}
U\chi_{G\bar{G}}(\vec{0})=1,
\label{PGC}
\end{align}
where $\chi_{G\bar{G}}(\vec{0})$ is the particle-particle bubble at zero
momentum and Matsubara frequency defined by
\begin{align}
\chi_{G\bar{G}}(\vec{0})\equiv -\frac{T}{{\cal N}_{\rm a}}\sum_{\vec{k}}G_\lambda(\vec{k})G_\lambda(-\vec{k}).
\label{chi_pp}
\end{align}

It follows from Eq.\ (32) and (33) of Ref.\ \onlinecite{Kita25} that 
$G_\lambda(\vec{k})$ can be expanded in $\bar\lambda$ as
\begin{align}
G_\lambda(\vec{k})=G(\vec{k})\left[1+\bar\lambda G(\vec{k})G(-\vec{k})+\cdots \right],
\label{G_lambda}
\end{align}
where $G(\vec{k})$ denotes Green's function for $\lambda=0$, and $\bar\lambda$ is defined by
\begin{align}
\bar\lambda\equiv \frac{{\cal N}_{\rm e}}{{\cal N}_{\rm a}}U \lambda.
\label{bLambda-def}
\end{align}
Using Eq.\ (\ref{G_lambda}), 
we can express the pseudogap condition (\ref{PGC}) for $T\lesssim T_{1{\rm PR}}$ 
up to the first order in $\bar\lambda$ as
\begin{align}
1=&\, -\frac{UT}{{\cal N}_{\rm a}}\sum_{\vec{k}}G(\vec{k})G(-\vec{k})
-2\bar\lambda \frac{UT}{{\cal N}_{\rm a}}\sum_{\vec{k}}\bigl[G(\vec{k})G(-\vec{k})\bigr]^2 .
\label{PGC2}
\end{align}
Let us subtract the corresponding equation at $T= T_{1{\rm PR}}$ with $\bar\lambda=0$
from Eq.\ (\ref{PGC2}).
We thereby obtain an expression of $\bar\lambda$ for $T\lesssim T_{1{\rm PR}}$ as
\begin{align}
\bar\lambda 
=&\,\frac{\displaystyle \frac{UT}{{\cal N}_{\rm a}}\sum_{\vec{k}}\left[G(\vec{k})G(-\vec{k})-G(\vec{k})G(-\vec{k})\Bigr|_{T=T_{1{\rm PR}}}\right]}{\displaystyle -2\frac{UT}{{\cal N}_{\rm a}}\sum_{\vec{k}}\bigl[G(\vec{k})G(-\vec{k})\bigr]^2} 
\equiv \frac{I_{\rm n}}{I_{\rm d}} .
\label{bLambda}
\end{align}

Both of its numerator $I_{\rm n}$ and denominator $I_{\rm d}$ can be evaluated analytically in a crude
approximation of using the non-interacting Green's function
$G_0(\vec{k})$ in place of $G(\vec{k})$.
Accordingly, we also replace $T_{1{\rm PR}}$ in Eq.\ (\ref{bLambda}) by the superconducting transition temperature 
$T_{{\rm c}0}$ of the mean-field theory given by Eq.\ (20) of Ref.\ \onlinecite{Kita25}.
Repeating the procedure of deriving the analytic expression for $T_{{\rm c}0}$ there,
we can transform the numerator of Eq.\ (\ref{bLambda}) in this approximation into
\begin{subequations}
\label{I_nI_d}
\begin{align}
I_{\rm n}\approx-UN(0)\ln \frac{T}{T_{{\rm c0}}},
\end{align}
where $N(0)$ is the density of states per spin and per site at the non-interacting Fermi energy $\varepsilon_{\rm F}^0$.
We can also transform the denominator in the present approximation as
\begin{align}
I_{\rm d}\approx &\, -\frac{2UT}{{\cal N}_{\rm a}}\sum_{\vec{k}}\frac{1}{(\varepsilon_n^2+\xi_{\bf k}^2)^2}
\notag \\
\approx &\, -2UTN(0)\sum_n  \int_{-\infty}^\infty \frac{d\xi_{\bf k}}{(\varepsilon_n^2+\xi_{\bf k}^2)^2}
\notag \\
\approx &\, -2UTN(0)\sum_n 2\pi i\frac{-2}{(2i\varepsilon_n)^3}=\frac{-2UN(0)}{(\pi T)^2}\frac{7}{8}\zeta(3)
\notag \\
\approx &\,\frac{-UN(0)}{T_{{\rm c}0}^2}\times 0.213,
\label{I_d}
\end{align}
\end{subequations}
where $\zeta(3)\approx 1.202$ is the Riemann $\zeta$ function,
and we have approximated (i) the density of states by the value at the Fermi energy and (ii) $T$ by $T_{{\rm c}0}$.
Substituting Eq.\ (\ref{I_nI_d}) in Eq.\ (\ref{bLambda}) with $T_{{\rm c}0}$ replaced by $T_{1{\rm PR}}$, we obtain
\begin{align}
\bar\lambda\approx -4.69T_{1{\rm PR}}(T_{1{\rm PR}}-T) .
\label{bLambda2}
\end{align}
The value of $\lambda$ at $T=0.098$ just below $T_{1{\rm PR}}=0.100$ in Fig.\ \ref{fig1} is $1.36\times 10^{-2}$,
whereas Eqs.\ (\ref{bLambda-def}) and (\ref{bLambda2}) yield $\lambda=7.82\times 10^{-3}$ 
for the same set of parameters $(T_{1{\rm PR}},T,\frac{{\cal N}_{\rm e}}{{\cal N}_{\rm a}}U)=(0.100,0.098,-0.12)$.
Thus, we realize that Eq.\ (\ref{bLambda2}) may be used as an order-of-magnitude estimate on how $\lambda$ develops initially.
The origin of the quantitative disagreement may be traced to (i) the complete omission of the self-energy in Green's function and (ii) the assumption of the constant density of states in transforming Eq.\ (\ref{I_d}).

\subsection{Discontinuity of ${\cal C}_V$}

The heat capacity ${\cal C}_V$ is obtained by performing the differentiation of Eq.\ (\ref{calC1}) in terms of Eq.\ (\ref{calE1}).
The discontinuity of  ${\cal C}_V$ at $T=T_{1{\rm PR}}$ in Fig.\ \ref{fig4} is caused by the linear emergence of
$\lambda$ in Fig.\ \ref{fig1}. 
To see this, we start by writing the relevant first-order variation of Eq.\ (\ref{calE1}) caused by the emergence of $\lambda$ as
\begin{align}
\delta {\cal E} =&\, 2T\sum_{\vec{k}}\biggl\{\left[\xi_{\bf k}+\mu+\frac{1}{2}\Sigma(\vec{k})\right]\delta G_\lambda(\vec{k})
\notag \\
&\, \hspace{10mm}+\frac{1}{2}G(\vec{k})\delta\Sigma_\lambda (\vec{k})\biggr\},
\label{deltaG1}
\end{align}
where we have used the fact that $\xi_{\bf k}+\mu\propto k^2$ is the kinetic energy independent of $\lambda$ and $\mu$.
The contribution of $\mu$ in Eq.\ (\ref{deltaG1}) can be omitted due to
the condition of the fixed electron number 
\begin{align}
2T\sum_{\vec{k}}\delta G_\lambda(\vec{k})=0 .
\end{align}
We subsequently express 
\begin{subequations}
\begin{align}
\delta G_\lambda(\vec{k}) =&\, -\bigl[G(\vec{k})\bigr]^2 \bigl[\delta \mu-\delta\Sigma_\lambda(\vec{k})\bigr],
\\
\frac{1}{2}=&\,\frac{1}{2}\bigl[i\varepsilon_n-\xi_{\bf k}-\Sigma(\vec{k})\bigr] G(\vec{k}),
\end{align}
\end{subequations}
based on Dyson's equation.
We can thereby transform Eq.\ (\ref{deltaG1}) into
\begin{align}
\delta {\cal E} =&\, 2T\sum_{\vec{k}}\biggl\{-\left[\xi_{\bf k}+\frac{1}{2}\Sigma_\lambda(\vec{k})\right]\bigl[G(\vec{k})\bigr]^2\delta\mu
\notag \\
&\,+\frac{i\varepsilon_n+\xi_{\bf k}}{2}\bigl[G(\vec{k})\bigr]^2\delta\Sigma_\lambda (\vec{k})\biggr\}.
\label{deltaG2}
\end{align}
We now approximate 
\begin{align}
\delta\Sigma_\lambda (\vec{k})\approx G(-\vec{k})\delta \bar\lambda,
\end{align}
based on Eq.\ (25) of Ref.\ \onlinecite{Kita25} by neglecting the vertex correction to $\delta\Sigma_\lambda$,
where $\bar\lambda$ is defined by Eq.\ (\ref{bLambda-def}).
We then replace $G$ by $G_0$ in Eq.\ (\ref{deltaG2}) and omit the self-energy contribution accordingly.
The resulting $\delta {\cal E}$ can be transformed as
\begin{align}
\delta {\cal E} \approx &\, 2T\sum_{\vec{k}}\biggl\{-\xi_{\bf k}\delta\mu
+\frac{i\varepsilon_n+\xi_{\bf k}}{2}G_0(-\vec{k})\delta\bar\lambda\biggr\}\bigl[G_0(\vec{k})\bigr]^2
\notag \\
=&\, 2\sum_{{\bf k}}\biggl(-\xi_{\bf k}\delta\mu
+\frac{1}{2}\delta\bar\lambda\biggr)\frac{\partial\bar{n}(\xi_{\bf k})}{\partial\xi_{\bf k}}
\notag \\
=&\, -{\cal N}_{\rm a} N(0) \delta\bar\lambda,
\label{deltaG3}
\end{align}
where $N(0)$ is the density of states per spin and per site at the Fermi energy, $\bar{n}(\xi_{\bf k})=(e^{\xi_{\bf k}/T}+1)^{-1}$ is the Fermi distribution function, and we have approximated
\begin{align}
\frac{\partial\bar{n}(\xi_{\bf k})}{\partial\xi_{\bf k}}\approx-\delta(\xi_{\bf k}),
\end{align}
as appropriate at low temperatures.
Substituting Eq.\ (\ref{bLambda2}) in Eq.\ (\ref{deltaG3}) and performing the differentiation of Eq.\ (\ref{calC1})
at $T=T_{1{\rm PR}}$, 
we obtain the discontinuity of the heat capacity as
\begin{align}
\varDelta {\cal C}_V\approx -{\cal N}_{\rm a} N(0)\frac{d\bar\lambda}{dT}\biggr|_{T=T_{1{\rm PR}}}\approx 
4.69 T_{1{\rm PR}}{\cal N}_{\rm a} N(0) .
\label{deltaC-analytic}
\end{align}
This value is about $0.713$ times the heat capacity
$${\cal C}_{V0}=\frac{2\pi^2}{3}T_{1{\rm PR}}{\cal N}_{\rm a} N(0)$$
of the corresponding non-interacting electron gas.
The discontinuity of $C_V$ at $T=T_{1{\rm PR}}$ in Fig.\ \ref{fig4} is $7.8\times 10^{-3}$,
whereas Eq.\ (\ref{deltaC-analytic}) yields $1.2\times 10^{-2}$ for the set of parameters
$(T_{1{\rm PR}},\frac{N(0)}{V_{\rm unit}})=(0.100,\frac{1}{4\pi^2})$.
Thus, Eq.\ (\ref{deltaC-analytic}) obtained through several crude approximations 
yields an order-of-magnitude estimate for the magnitude of the discontinuity.

\end{document}